\documentstyle[prl,aps,multicol]{revtex}  
\begin{document}
\newcommand{\be}{\begin{equation}}
\newcommand{\ee}{\end{equation}}
\newcommand{\bq}{\begin{eqnarray}}
\newcommand{\eq}{\end{eqnarray}}
\def\id{{\rm 1\kern-.34em 1}}
\def\RR{{\rm I\kern-.20em R}}

%
\title{Algebraic equivalence between  certain models for 
superfluid--insulator transition 
}
\author{Luigi Amico}
\address{Dipartimento di Metodologie Fisiche e Chimiche (DMFCI), 
Universit\'a di Catania, viale A. Doria 6, I-95125 Catania, Italy\\
Istituto Nazionale per la Fisica della Materia, Unit\'a  di Catania, Italy} 
\maketitle
\begin{abstract}
Algebraic  contraction 
is proposed to realize mappings between  models Hamiltonians.
This  transformation contracts  the
algebra of the  degrees of freedom underlying  the Hamiltonian. 
The rigorous mapping between the anisotropic $XXZ$ Heisenberg model, 
the Quantum Phase Model, and the Bose Hubbard Model is established  
as the  contractions of the algebra $u(2)$ underlying the 
dynamics of the $XXZ$ Heisenberg model. 
\end{abstract}
%
The problem of mapping between not equivalent 
algebras was solved, in mathematical physics, years ago by 
In\"on\"u and Wigner~\cite{INONU-WIGNER} and subsequently generalized 
by Saletan~\cite{SALETAN} when they founded the concept of 
contraction of a Lie algebra ${\cal A}$~\cite{CONTRACTION}.
Algebraic contraction is a transformation 
which may be singular on  ${\cal A}$'s basis 
(namely, the  kernel of the transformation is non trivial), 
while it is regular on its commutation brackets~\cite{NOTE-TENSOR}.     
Applications  of   
algebraic contractions in condensed matter physics trace back to studies of    
Umezawa and coworkers~\cite{UMEZAWA}. They  shown, under quite general 
hypothesis, that in a zero temperature phase transition, 
the symmetry of the system in the 
disordered phase (is {\it rearranged}) contracts
(through contraction of the algebra spanned by  degrees of freedom of 
the system), onto the symmetry of the ordered phase.
For example, in Heisenberg ferromagnets the broken symmetry 
$so(3)$ (which is the spin algebra and which  accounts for 
the rotation symmetry of the magnetization in the paramagnetic phase)
is contracted onto the euclidean symmetry  $e(2)$ of the traslators
(which accounts  for  the traslational invariance 
and for  the rotation symmetry 
around the magnetization axes of the ordered state)~\cite{UMEZAWA}. 

In the present work, I  apply contractions in physics 
toward a slightly different direction. That is: 
Contractions of algebras spanned by
the degrees of freedom of the system as    
establishing a link between  models which are intrinsically 
distinct (in the sense they are not unitarly connectible).
I will provide an application of 
such idea  in condensed matter physics: I will show that contractions  
can provide  exact mapping between the Bose  Hubbard 
model, the quantum Josephson model and certain anisotropic Heisenberg 
model. The motivation is to found rigorously 
the relation between  these three models,  
which is  employed (using physical grounds) to describe low temperature 
behaviour 
in  various mesoscopic systems characterized by superfluidity 
(for applications of theses three models in mesoscopic 
physics, see for istance Refs.~\cite{BHM,QPM}) 
The  Bose Hubbard Model 
(BHM)  describes a lattice gas of interacting charged bosons.
Its elementary degrees of freedom are the bosonic  
site $j$ annihilation $a_j$, creation $a_j^\dagger$, and number $n_j$ operators.
The  Quantum Phase Model (QPM) is 
largely employed in the physics of Josephson junctions arrays
since it can describe the competition between quantum phase 
coherence and Coulomb blockade.
The elementary 
degrees of freedom entering the QPM are the phases of the superconducting 
order parameter $\phi_j$ and the charge unbalance to charge neutrality 
$N_j := -i \partial_{\phi_j}$ 
(its eigenvalues range in $(-\infty, +\infty )$) in the island $j$.
These two variables are considered as 
canonically conjugated in the QPM.
\\
The phase diagrams of BHM and QPM were analyzed by many 
authors~\cite{PHASE-DIAGRAM}.
They describe zero temperature  quantum phase transitions
between  incompressible insulators and  coherent superfluid phases.
\\
Finally, the  $XXZ$ anisotropic Heisenberg 
model~\cite{IUZUMOV} shows a 
low temperature behaviour 
related to those ones of BHM and QPM. In particular, its zero temperature 
phase diagram shows phase transitions from  paramagnetic to  canted 
phases that can be interpreted as insulator to superfluid 
phase transition~\cite{XXZ}.
\\
Up to the present study, the relation between the BHM, the QPM and 
the XXZ model consisted in the fact that they belong to the same universality 
class.
Unitary transformations mapping one model on each other  do not  exists.
In fact, the arguments usually employed to relate such models on each other 
did not want to be rigorous~\cite{PHASE-DIAGRAM}. For instance, 
the phase--number variables entering the QPM 
cannot be thought as mathematically originated from bosonic 
operators in BHM since a {\it no--go} theorem forbids 
$ a_j \sim \sqrt{ n_j } e^{i \phi_j }$,
$a^\dagger_j \sim e^{-i\phi_j}\sqrt{n_j}$ 
(``even with the widest reasonable latitude of 
interpretation''~\cite{CARRUTHERS,DUBIN})
as long as the phases $\phi_j$ are hermitian and 
canonically conjugated to a 
bounded  (from below) $n_i$ (as it is the bosonic number operator).
A way out from this difficulty is realized in QPM by removing the  
hypothesis of  boundness from below of $n_i$. 
It is worthwhile noting that connections between $n_i$ and $N_i$ 
cannot be  unitary since their spectra are not isomorphic.
Even more, such two operators cannot be unitarly connected to spin operators
since unitary transformations cannot transform bounded into unbounded 
operators.
In contrast,  algebraic contractions can do such a job.
I will use it  as the 
crucial tool to realize the mapping between the three models I deal with.
Such a transformation induces  also the mapping of the 
matrix elements of the Hamiltonians as well of the phase boundaries 
in their phase diagrams. Following this direction, contraction was employed 
in  
Ref.~\cite{AMPE} to  
map the zero temperature phase diagram 
of the BHM onto the phase diagrams of the QPM and of 
the XXZ model within (suitable) mean field approximation.   

The paper is organized as follow. After having outlined  the general procedure 
consisting in  contracting the underlying algebra characterizing quite 
general Hamiltonians,  then it is applied to realize the 
mapping between  the BHM, the QPM, 
and the $XXZ$.

I assume models Hamiltonian  on a lattice $\Lambda$ writable  in 
terms of generators of a given Lie algebra ${\cal A}=\oplus_{i\in \Lambda} 
\; g_i$ having the form:  
\begin{eqnarray}
\label{GENH}
H=
\sum_{v,w} && \sum_{i,j}h_{v,i}\xi_{i,j} h_{w,j} - \\
&&\sum_{\alpha\neq \alpha'} \sum_{i,j} \left (e_{\alpha,i} 
\zeta_{ij} e_{\alpha',j}+ 
e_{\alpha',j} \zeta_{ji} e_{\alpha,i} \right ) \, , \nonumber 
\end{eqnarray}
where $\xi_{i,j}$ and $\zeta_{ij}$ are real parameters. The local algebra 
$g_i$ is defined as $g_i:= \id \otimes \dots \otimes \id\otimes g \otimes\id \otimes \dots\otimes \id$ with  $g$
at the $i$--th lattice position; the sum on $\alpha$'s runs on 
the set of simple roots of $g_i$; $i,j \in \Lambda$.  
Any $g$  is assumed a rank--$r$ semisimple Lie algebra of dimension $dim(g)$~\cite{NOTE} whose generators, 
in the Cartan--Weyl 
normalization, obey the standard commutation rules: 
$[h_{v,i},h_{w,i}]=0$, $(v,w=1\dots r)$, 
$[h_{v,i}, e_{\alpha,i}]=\alpha_v e_{\alpha,i}$, 
$[e_{\alpha,i},e_{\beta,i}]=c_{\alpha,\beta}^{\alpha+\beta} e_{\alpha+\beta,i}$ if 
$\alpha +\beta \neq 0$ and 
$[e_{\alpha,i}e_{-\alpha,i}]=\alpha^v h_{v,i}$; 
$c_{\alpha,\beta}^{\gamma}$ are the structure constants 
(the sum convention is assumed). 
For the global algebra ${\cal A}$, contractions can be done as products
of local contractions  
of each $g_i$~\cite{NEC&SUFF-COND}. That is,  
as transformation $R=R(\epsilon;p):=\prod_i R_i(\epsilon;p)$ where 
$R_i(\epsilon;p):=\id \otimes \dots \otimes \id \otimes r(\epsilon;p)\otimes \id 
\otimes \dots \otimes \id$ (where $\epsilon$ is a real variable, and 
$p$ is a real parameter). The matrix $R_i(\epsilon;p)$ maps $g_i$
onto another algebra $g_i'$ which is in one to one  correspondence with $g_i$ 
when $\epsilon \neq 0$; additionally, there 
exist the limit $\epsilon \rightarrow  0$ for any value of the parameter $p$:
\begin{eqnarray}
&&\lim_{\epsilon \rightarrow 0}[h'_{v,i},h'_{w,i}]=0 \;, \nonumber \\  
&& \lim_{\epsilon \rightarrow 0} [h_{v,i}', e_{\alpha,i}']=\alpha(\epsilon;p)_v e_{\alpha,i}' \; , \nonumber \\ 
&&\lim_{\epsilon \rightarrow 0}
[e_{\alpha,i}',e_{\beta,i}']=c(\epsilon;p)_{\alpha,\beta}^{\alpha+\beta} e_{\alpha+\beta,i}' \quad \alpha +\beta \neq 0 \nonumber \\ 
&& \lim_{\epsilon \rightarrow 0} 
[e'_{\alpha,i}e'_{-\alpha,i}]=\alpha^v(\epsilon;p) h_{v,i}' \, . 
\end{eqnarray}
The operators: $h_{v,i}':=r^v(\epsilon ;p)h_{v,i}$ and 
$e_{\alpha,i}':=r^\alpha(\epsilon;p)e_{\alpha,i}$, where $r^v(\epsilon ;p)$ and $r^\alpha(\epsilon ;p)$ are submatrices of 
$r(\epsilon ;p)$ acting only on Cartan subalgebra (spanned by the set of 
$h_{v,i}$, $v\in (1\dots r)$)
and root space  separately, (spanned by the set of 
$e_{\alpha,i}$, $\alpha \in (1\dots dim(g)-r)$)
define the transformed basis of the new algebra $g_i':= R_i g_i$ 
($dim(g_i')\equiv dim(g_i)$).  
The algebra  ${\cal A}':= R[{\cal A}]$ which may be not unitarly 
equivalent to ${\cal A}$, 
is the contraction of ${\cal A}$. 
\\
As result of the contraction $R[{\cal A}]$,
the Hamiltonian~(\ref{GENH}) is transformed into the contracted Hamiltonian as
\begin{eqnarray}
\label{RGENH}
H\rightarrow H':=R H= \sum_{v,w} && \sum_{i,j}h_{v,i}'\xi_{i,j}h_{w,j}' - \\
&&\sum_{\alpha\neq \alpha'} \sum_{i,j} \left (e_{\alpha,i}' 
\zeta_{ij}  e_{\alpha',j}'+h.c \right )\, , \nonumber 
\end{eqnarray}  
where the set of new  degrees of freedom $\{ h_{v,i}'\, , \,e_{\alpha,i}' \} $
does not have to  be unitarly equivalent to the set of the original variables 
$\{ h_{v,i} \, , \,e_{\alpha,i} \} $. 
 
Now I apply the scheme developed above to map the 
$XXZ$ model on to BHM, and  QPM~\cite{STABILITY}. 
In this case,  it is sufficient to 
consider the algebra $g_i$ having  rank--$1$;
thus the sum on simple roots in Hamiltonian~(\ref{GENH}) reduces to a single term coupling the positive with the negative root operators. 
The Hamiltonian~(\ref{RGENH}) becomes:
\begin{eqnarray}
\label{GENH-rank1}
H'=
\rho \sum_{i} h_{i}' +&&
\sum_{i,j}h_{i}'\xi_{ij} h_{j}' - \\ 
&&\sum_{\langle i,j\rangle} \left (e'_{+,i} 
\zeta_{i,j} e'_{-,j}+ 
e'_{+,j} \zeta_{j,i} e'_{-,i} \right ) \; , \nonumber 
\end{eqnarray}
where  the linear term in Cartan generators has been isolated; 
the sum of the roots operators $e_{\pm \alpha,i}'\equiv  e'_{\pm, i}$ 
involves only nearest neighbouring site indices.
\\
The algebra $g_i'$ is to be taken as  the rotated 
$R_i(\epsilon,p) [u(2)_i]$ of $u(2)_i=\id \oplus su(2)_i$ which is characterized  by   
$
[ J^{3}_i,J^{\pm}_j ]=\pm \delta_{i,j} J^{\pm}_i$, 
$ [ J^{+}_i,J^{-}_j ]=2 \delta_{i,j} J^{3}_i $,
$ [ {\bf J}_i,\id  ]=0 $
with standard representations 
$
J^{3}_i|J_i,m_i\rangle =m_i|J_i,m_i\rangle $ 
$ 
J^{\pm}_i|J_i,m_i\rangle =\left [(J_i\mp m_i) (J_i\pm m_i+1)\right ]^{1/2}
|J_i, m_i \pm 1 \rangle \,$.
The matrix $R_i(\epsilon,p)$  defines the change of  ``basis'' of $u(2)_i$ 
(as vector space)   
$(e'_{+,i}\; , \; e'_{-,i}\; , \; h'_i\; , \; \id)^T= r(\epsilon;p)
( J^{+}_i\;,\; J^{-}_i \; ,  \;J^{3}_i \; ,\; \id)^T$ with:
\begin{equation}
r(\epsilon;p):= \left(
\begin{array}{clcr}
\epsilon & 0 & 0 & 0 \\
0 & \epsilon & 0 & 0\\
0 & 0 & 1 & {{p}\over{2 \epsilon^2}} \\
0 & 0 & 0 & 1 
\end{array}
\right) \, .
\label{R-CONTR}
\end{equation}
The generators ${\bf h}'_i:= \{ h'_i\;, \; e'_{\pm,_i} \}$ 
are expressed in terms of ${\bf J}_i$ as 
\begin{equation}
e'_{\pm,i} =\epsilon J^{\pm}_i \makebox{\hspace{1cm}} h'_i=J^3_i+
\id {{p}\over{2 \epsilon^2}} \,.
\label{HVSJ}
\end{equation}
The commutation rules of $g_i$ are: 
\begin{eqnarray}
[ h'_i,e'_{\pm,j} ] &=&\pm \delta_{i,j} e'_{\pm,i} \\ \nonumber
[ e'_{+,i},e'_{-,j} ] &=&\delta_{i,j} (2 \epsilon^2 h'_i -p \id )
 \\ \nonumber
 [ {\bf h}'_i,\id  ]&=&0 \; . 
\label{COMMUCONTR}
\end{eqnarray}
The matrix elements of the  Hamiltonian~(\ref{GENH-rank1}) are:
\begin{eqnarray}
\label{MATRIX-XXZ}
\hspace{-2cm}\langle J',m'|H'|J,m\rangle = 
\rho \sum_{i} B_i\delta_{m_i',m_i} 
&&+ \sum_{i,j}B_i\xi_{ij} 
B_j\delta_{m_i',m_i}
\delta_{m_j',m_j} \\  
&&-\sum_{\langle i,j\rangle}  \left (\zeta_{i,j}C_{i,j}
\delta_{m_i',m_i+1}\delta_{m_j',m_j-1} + 
i \leftrightarrow j\right ) \; , \nonumber 
\end{eqnarray}
\noindent
where $|J,m \rangle:= \otimes_i |J_i,m_i\rangle$,  
$B_i:= m_i +{{p}\over{2 \epsilon^2}}$ and: 
\\
$C_{i,j}:=  
\epsilon^2 \sqrt{(m_i-J_i)(m_j+J_j)
(m_i+J_i+1)(m_j-J_j+1)}$.
A trivial case corresponds to leaving  $\epsilon$ as finite and 
setting $p=0$. 
In such a case, $\epsilon$ can be normalized; $r(\epsilon;p)$ is 
isomorphic to the identity:
$(e'_{\pm,i}\; ,\; h'_i\; , \; \id )\equiv 
( J^{\pm}_i\;, \;J^{3}_i \; ,\; \id )$.
Thus, the resulting Hamiltonian~(\ref{GENH-rank1}) 
is the $XXZ$ model  
where
$\rho$, $\xi_{i,j}$, $\zeta_{i,j}$ can 
be interpreted as the external magnetic field and the magnetic exchange 
coupling constants respectively.
\\
Instead, the contraction of $\oplus_i u(2)_i$  
is realized  through  the  limit $\epsilon \rightarrow 0$:
The transformation $R$ is singular, 
but the commutation rules~(\ref{COMMUCONTR}) are well defined. 
\\
There are two possible choices: 
{\it i)}: $\epsilon \rightarrow 0$, $p=0$;
{\it ii)}: $\epsilon \rightarrow 0$,  $p \neq 0$.
\\
In the case {\it i)} the  commutation rules~(\ref{COMMUCONTR}) contract to: 
\begin{eqnarray}
[ h'_{i},e'_{\pm,j} ]=\pm &&\delta_{i,j} e'_{\pm,_i} \;,\;
 [ e'_{+,i},e'_{-,j} ]=0 \nonumber \\
&& [ {\bf h}'_i ,\id ]=0 \; .
\label{E2}
\end{eqnarray}
Such commutation relations are isomorphic to the commutation relations 
of the algebra $e(2)_i\oplus \RR$. Thus,  
the Hamiltonian~(\ref{GENH-rank1}) contracts to the QPM:
$
H_{_{QP}} =  \sum_{i ,j} (N_i-N_x) \, V_{ij} \, (N_j-N_x)
-E_{_J}/2 \sum_{< i , j >} \cos (\phi_i-\phi_j  )
$,  
where $N_x \sum_j V_{i,j}\equiv \rho\; \forall i$, $V_{i,j}\equiv \xi_{i,j}$,
and $\delta_{<i,j>} E_J/2 \equiv \zeta_{i,j}$.
Where $[N_i,\phi_j]=i \delta_{i,j}$. Such a commutation relation induces
$ [ N_{i},e^{\pm i \phi_j} ]=\pm \delta_{i,j}  e^{ \pm i \phi_j}$;
from the hermitianity of $\phi_j$ it comes also that: 
$ [e^{+i \phi_i},e^{-i  \phi_j} ]=0$ (compare with~(\ref{E2})).
\\
The representations of the contracted algebra~\cite{CONTRACTION,CELEGHINI} $e(2)_i$ are  the contraction 
of the representations of $u(2)_i$ for large $J_i$:
$
\langle J_i,m_i |\epsilon J^{\pm}_i  |J_i,m_i'\rangle \rightarrow l_i 
\delta_{m_i',m_i\pm 1}
$
requiring that 
$\epsilon J_i\rightarrow l_i $, $l_i=\epsilon J_i$ being finite real numbers; 
whereas $\langle J_i, m_i|J^{3}_i|J_i,m_i'\rangle
\rightarrow N_i\delta_{m_i',m_i}$ whose eigenvalues  can range 
in $(-\infty, +\infty )$ after having done the limit 
$J_i\rightarrow \infty$.  
In fact, this contraction (of representations) can be seen as 
suitable large $J$ ($J_i\equiv J \;, \forall i$) 
limit of the    Villain 
realization of spin algebra~\cite{AMPE,NOTESC} 
$J^+_j:= e^{i\phi_j}\sqrt{(J+1/2)^2-(J^3_j+1/2)^2}$, $J^-_j=(J^+_j)^\dagger$
where $J^3_j$ fulfills 
$[J^3_j,e^{\pm i\phi_l}]=\pm \delta_{j,l}e^{\pm i\phi_l}$. In the Ref.~\cite{AMPE}
it is shown that $J$ plays the role of the Cooper pairs density in the 
islands.
\\
The matrix elements of the contracted Hamiltonian   can be 
obtained through $B_i\equiv N_i$ and 
$C_{i,j}\rightarrow l_i l_j\sqrt{1-(N_i N_j/l_i l_j)^2 \epsilon^2}$ 
in~(\ref{MATRIX-XXZ}). 
\\
In the case {\it ii)}, 
$p$ can be normalized. The algebra resulting from the contraction    
of (\ref{COMMUCONTR}) is:
\begin{eqnarray}
 [ h'_{i},e'_{\pm,j} ]=\pm &&\delta_{i,j} e'_{\pm,_i} , 
 [ e'_{+,i},e'_{-,j} ]=\delta_{i,j} \id  \nonumber \\
&& [ {\bf h}'_i, \id ]=0 \, .
\label{H4}
\end{eqnarray}
Such commutations are isomorphic to the ``single boson algebra''  
$(h_4)_i\oplus \RR$: 
spanned by operators $n_{i}$, $a^{\dagger}_{i}$ and $a_{i}$ 
fulfilling 
$[ n_{i},a_j]=-\delta_{i,j} a_i$,
$ [ n_{i},a_j^\dagger  ]=\delta_{i,j} a^\dagger_i$, 
$ [a_i, a_j^\dagger  ]=\delta_{i,j}$ (compare with~(\ref{H4})).
This set of operators are the microscopic operators of the BHM:
$
H_{_{BH}}=-\mu \sum_i n_i
+\sum_{i,j} n_{i}U_{i,j}  n_{j}
-\sum_{\langle i,j\rangle} (a^{\dagger}_{i}t_{j,i}
a_{j}+ a^{\dagger}_{j} t_{i,j} a_{i} ) 
$,
on which Hamiltonian~(\ref{GENH-rank1})
is contracted ($\mu\equiv -\rho$, $U_{i,j}\equiv \xi_{i,j}$,
and $t_{i,j} \equiv \zeta_{i,j}$).
\\
The representations of the contracted algebra~(\ref{H4}) are
$ 
\langle J_i, m_i|\epsilon  J^{\pm}_i|J_i,m_i'\rangle
\rightarrow 
\sqrt{n_i+1/2(1\pm 1)}  \delta_{n_i^{'},n_i\pm 1}
$ 
where  $J_i+m_i\rightarrow n_i$ (keeped finite in the limit)
and  $2J_i\epsilon ^2\rightarrow 1$ for  
$J_i\rightarrow \infty$, $m_i\rightarrow -\infty$;
whereas
$
\langle J_i, m_i| J^{3}_i+1/(2 \epsilon^2)\id |J_i,m_i'\rangle 
\rightarrow n_i\delta_{m_i',m_i}
$. I point out that the matrix elements of the bosonic number 
operator are obtained renormalizing angular momentum's 
matrix elements by $1/\epsilon^2 \rightarrow \infty$ 
since $m_i$, originally ranging in $(-J_i\dots J_i)$, must cover the interval
$(0\dots \infty)$. 
In fact,  $1/\epsilon^2 \sim J$; then, this contraction (of representations) 
can be seen as 
suitable large $J$ limit of the spin algebra in the  Holstein Primakoff 
realization~\cite{AMPE,NOTESC}: 
$J^+_j:= \sqrt{2 J} a^\dagger_j\sqrt{1-n_j/(2J)}$, $J^-_j=(J^+_j)^\dagger$,
$J^3_j:=n_j-J$. 
In the Ref.~\cite{AMPE}
it is shown that $J$ can be interpreted as the bosons density.
\\
The matrix elements of the contracted Hamiltonian   can be 
obtained through $B_i\equiv n_i$ and 
$C_{i,j}\rightarrow \sqrt{(n_i+1)n_j}
\sqrt{1-\epsilon^2}$ in~(\ref{MATRIX-XXZ}).
 
The algebra $(h_4)_i$ can be contracted further.
Such a contraction induces the mapping between the BHM and the QPM as 
follow.
\\
The BHM Hamiltonian can be written trivially as 
Hamiltonian~(\ref{GENH-rank1}), whose algebra is the enveloping of $g_i$
spanned by the transformed $R_i(\epsilon,2 p) [(h_4)_i]$, for $p=0$, $\epsilon=1$. 
For generic $\epsilon$,
$g_i$ is spanned by the operators
$(A^{+}_i\; , \; A^{-}_i\; ,\; A^3_i\; , \; \id )^T= 
r(\epsilon;p) ( a^{\dagger}_i\;,\; a_i \; ,  \; n_i \; ,\; \id )^T $. 
The new generators ${\bf A}_i:=\{ A^{\pm}_i\;, \; A^{3}_i\}$  are expressed in terms of 
$\{a^\dagger_i \; , \; a_i \;,\; n_i\}$  as
\begin{equation}
A^+_i=\epsilon  a^\dagger_i \; ,\; A^-_i=\epsilon  a_i\; , \; A^3_i=n_i+{{p}\over{ \epsilon ^2}} \, ,
\label{AVSa}
\end{equation}
whose commutation rules  are 
\begin{eqnarray}
 [ A^{3}_i,A^{\pm}_j ]=\pm \delta_{i,j} A^{\pm}_i \; &,& \; [ A^{+}_i,A^{-}_j ]= \epsilon^2 \id \\ \nonumber 
\hbox{\hspace{2cm}} [ {\bf A}_i,\id ]&=& 0 \,.
\label{COMMUCONTRBH}
\end{eqnarray}
The limit  $\epsilon \rightarrow 0$ (with finite $p$) realizes the (local) contraction of $(h_4)_i \oplus \RR$ in $e(2)_i\oplus \RR$ and thus it induces  the contraction of the 
underlying  algebra of the  BHM on the QPM's one.
\\
I point out that since the generators ${\bf A_i}$ can be seen as 
contraction of the vectors ${\bf J_i}$, the QPM is recovered 
as ``first order'' contraction of the BHM but also 
as a ``second order'' contraction of the XXZ model. 
This implies, in particular, that the coupling constants of the BHM and 
the QPM are related as: $E_J\simeq \epsilon t$. This suggests 
how superfluidity should be enhanced in  the 
BHM respect to the QPM.
 
In conclusion, the contractions of the algebra 
${\cal A}=\oplus_i u(2)_i$ underlying XXZ model,
realize the exact mapping between the BHM, QPM and XXZ model.
Using representations of ${\cal A}$, this was already employed 
in the Ref.~\cite{AMPE}
to relate  the zero temperature 
phase diagram of the  XXZ model, with those ones of the 
BHM and  QPM. In particular, identifyng the mapping between these three 
models was crucial to bypass the problem  
concerning the  coherent state representation of the phase--number algebra 
which is the basic difficulty involved in the semiclassical 
representation of the QPM.
\\
As noted by Umezawa~\cite{UMEZAWA}, such a contraction limit corresponds 
to consider low energy physical regime of the spin problem. 
Thus, the BHM and the QPM can be considered as low energy 
effective descriptions of the XXZ model. 
\\
I point out that 
the algebras underlying the three models as well their spectra
are left distinct by the transformation above 
since the latter is, in particular, not  unitary. 
This feature  should  be 
considered positively since mappings based on contractions 
can connect distinct physical scenarios.  
As it was already noted in Ref.~\cite{DAS,JOSE}, for istance, 
the difference 
between  the Casimir operators of $e(2)$ and of spin algebras 
motivates qualitative difference  between QPM's and XXZ model's 
zero temperature phase diagrams: 
in the QPM one's  a metallic phase  can exist; in XXZ one's 
such a metallic  phase cannot exist.    
In this sense, mappings based on contractions express 
relations which are ``weaker'' than those ones 
based on unitary mappings. 
\\ 
Mappings based on contractions can be applied toward two different directions.
First, they might serve to group the set of all mutually contracted models
in ``equivalence classes'' following the same procedure known 
in group theory. There, the classification of algebras
was considerably simplified by contractions which reduce 
the number  of eventually independent algebras~\cite{CONTRACTION}. 
In the same way, properties of models in the same equivalence class can 
be stated succinctly and perspicaciously, analoglously to what is done having
introduced the concept of universality class 
in the theory of phase transition~\cite{AMIT}.
\\  
Finally, contractions  might  be applied 
to the theory of integrable systems: 
properties of integrable models might be related 
to properties corresponding to non integrable models. 
It is whorthwhile noting that the same procedure could not be 
persecuted through unitary relations   
since the latter can connect  properties of integrable models to corresponding 
properties  of  models which are still integrable. 
In particular,  exact  properties of one dimensional   
QPM and  BHM (which resist to be exactly solved) might  be argued 
from corresponding properties of the XXZ model which, instead, 
is integrable in one dimension. Work is in progress along this direction. 

\bigskip
I would like to thank U. Eckern,  G. Falci, R. Fazio, G. Giaquinta, 
A. Osterloh, V. Penna,  
M. Rasetti, and J. Siewert for constant support and suggestions.

\vspace{1cm}


\begin{thebibliography}{100}
\bibitem{NOTE-TENSOR}All the properties concerning algebra's 
	linear automorphisms   are totally characterized 
	by algebra's constants structure 
	which  behave like tensors under  unitary transformations. 
	Since contractions are not  unitary transformation,
	they relate not equivalent algebras.
\bibitem{NEC&SUFF-COND} For necessary and sufficient conditions must fulfil
	a Lie algebra ${\cal A}$ to be contractible see~\cite{CONTRACTION}
\bibitem{STABILITY} The algebraic contractions studied in the following 
	are the contractions respect to the stability subalgebra 
	of $u(2)$.
\bibitem{NOTESC} It is worthwhile noting that such large $J$ limit is not 	
	the (semi)classical limit since $m$ has to be considered 
	comparable with $J$. Contractions of representations  
	can be realized through 
	Holstein--Primakoff or Villain realizations of the spin algebra.
	Instead, in the present paper contractions are realized algebraically.
\end{thebibliography}
\end{document}